\documentclass[aps,prb,showpacs,preprintnumbers,amsmath,amssymb,superscriptaddress]{revtex4}
\usepackage{dcolumn}
\usepackage{bm}
\usepackage{amsmath}
\usepackage{feynmp}

\usepackage{graphicx}

\begin{document}

\title{Phase Structure of the Topological Anderson Insulator}

\author{ Dongwei Xu}
\affiliation{Department of Physics, Oklahoma State University, Stillwater, Oklahoma 74078, USA}
\author{Junjie Qi}
\affiliation{Institute of Physics, Chinese Academy of Sciences, Beijing 100190, China}
\author{Jie Liu}
\affiliation{Department of Physics, Hong Kong University of Science and Technology, Clear Water Bay, Hong Kong, China}
\author{Vincent Sacksteder IV}
\email{vincent@sacksteder.com}
\affiliation{Institute of Physics, Chinese Academy of Sciences, Beijing 100190, China}
\author{X. C. Xie}
\affiliation{International Center for Quantum Materials, Peking University, Beijing 100871, China}
\author{Hua Jiang}
\affiliation{International Center for Quantum Materials, Peking University, Beijing 100871, China}


\date{\today}

\begin{abstract}
We study the disordered topological Anderson insulator in a 2-D (square not strip) geometry.  We first report the phase diagram of finite systems and then study the evolution of phase boundaries  when the system size is increased to a very large $1120 \times 1120$ area.  We establish that  conductance quantization can occur without a bulk band gap, and that  there are two distinct scaling regions with quantized conductance: TAI-I with a bulk band gap, and TAI-II with localized bulk states.        We show that there is no intervening insulating phase between the bulk conduction phase and the TAI-I and TAI-II scaling regions, and that there is no metallic phase at the transition between the  quantized and insulating phases. Centered near the quantized-insulating transition there are very broad peaks in the eigenstate size and fractal dimension $d_2$; in a large portion of the conductance plateau eigenstates grow when the disorder strength is increased. The fractal dimension at the peak maximum is $d_2 \approx 1.5$.   Effective medium theory (CPA, SCBA)  predicts well  the boundaries and interior of the gapped TAI-I scaling region, but fails to predict all boundaries save one of the ungapped TAI-II scaling region.  We report conductance distributions near several phase transitions and compare them with  critical conductance distributions for well-known models.

\end{abstract}

\maketitle

Topological insulators  \cite{Fu07, Zhang09, Hasan10, Qi10, Liu10b, Jiang09} are the focus of considerable interest for their quantized edge conductance which is similar to that seen in the integer quantum hall (IQH) effect.  Unlike IQH systems, TIs require neither magnetic fields nor low temperatures.    TIs exhibit a gap in the spectrum of bulk states, and bridging that gap is a band of edge states; if the Fermi energy is within the gap then electrons flow only  along the edge and not in the bulk.  In two dimensions there are only two  edge states, and they are helical, which means that they have opposite spins and opposite momenta.  If spin is conserved then backscattering is prohibited and the edge conductance is quantized.    Therefore the bulk band gap coincides with a plateau of quantized conductance.    All these properties are consequences of bulk band inversion in a system with time-reversal ($T$) symmetry, and are topologically protected against small perturbations which are $T$-symmetric, such as weak non-magnetic disorder.  

Li et al \cite{Li09} studied the Bernevig-Hughes-Zhang (BHZ) Hamiltonian \cite{Bernevig06a} which models the  2-D topological insulator  $HgTe$, but they tuned the mass parameter to a positive value ($M = 1$ meV) so that there is no inversion of bulk bands, no  band of edge states,  and no topological physics.  Li et al. introduced non-magnetic disorder and found a region of quantized conductance at intermediate disorder strengths, which implies the existence of conducting edge states.  Since their $M=1$ meV model has no edge states at zero disorder, they believed that the observed edge states are not caused by the band inversion which is central to topological insulators.  Moreover, Li et al. believed that the quantized region occurs in  a region where the bulk states are localized, not in a band gap like topological insulators.  They dubbed the observed quantized region a "topological Anderson insulator phase".  

 Soon afterward Groth et al \cite{Groth09} published a paper titled "Theory of the Topological Anderson Insulator"  which argued that  the topological Anderson insulator can be explained by mass inversion, the same as the zero-disorder topological insulator.  They used the Coherent Potential approximation (CPA, or SCBA) \cite{Soven67, Taylor67} to calculate an effective Hamiltonian which includes some disorder effects.    Renormalized strengths of the mass term and Fermi level can be extracted from the effective Hamiltonian.   Strong enough disorder changes the sign of the mass term, producing a band inversion and  topologically protected edge states.    Yamakage et al \cite{Yamakage11} and Prodan \cite{Prodan11} have performed numerical calculations of the evolution of the TAI conductance plateau when both the mass and the disorder strength are varied separately.  They studied cases which were intermediate between an ordinary TI (negative mass, no disorder) and a TAI (positive mass, finite disorder), and showed that the TI conductance plateau changes continuously into the TAI conductance plateau.  This numerical evidence indicates that the TAI conductance plateau is caused by inversion of the bulk bands.  

Groth et al \cite{Groth09}  argued that the weak-disorder boundary of the conductance plateau is a crossing from the bulk band into a band gap, and is not an Anderson transition. This implies that the absence of bulk conductance near this boundary is due to an absence of bulk states rather than their localization.  Groth et al also gave numerical evidence that the weak-disorder boundary of the conductance plateau matches well to the CPA's prediction of the bulk band edge, and that the conductance plateau is accompanied by a bulk band gap.  All of these findings were qualified by a proviso restricting their validity to "small systems accessible to numerical calculations"  \cite{Groth09}.   

Recently Chen et al \cite{Chen11} and Zhang et al \cite{Zhang11b} have found evidence that the TAI conductance plateau is populated by bulk states.  Moreover,  Guo et al \cite{Guo10} and Chen et al \cite{Chen11} plotted the region where the CPA predicts a bulk gap, and demonstrated that this region is much smaller than the conductance plateau which has been determined numerically.   Outside the gapped region Chen et al \cite{Chen11} found numerical evidence that  robust edge states coexist with bulk states, and they also  reported a metallic phase lying between the quantized and insulating phases.  These results  reveal  ongoing questions about the precise nature of the TAI conductance plateau and about the overall phase diagram.

In disordered systems there are two important length scales: the scattering length, and the localization length which is typically much longer.  The  scattering length scale regulates the density of states, while the localization length scale regulates conduction and localization.  The CPA correctly includes scattering physics but discards the quantum interference processes which determine the localization length scale.   Although the CPA does correctly predict the density of states because this quantity is controlled at leading order by scattering physics, it is unable to predict the conductivity of systems that are bigger than the localization length.  This was already obvious in the original TAI papers, which show a quantized-to-insulator transition at large disorder that is not predicted by the CPA.   The TAI phase diagram could change substantially in the infinite volume limit. 

In this light it is important to note that all studies of the TAI have been limited to small system sizes.  Studies of the conductance have minimized the computational time by examining pseudo-one-dimensional rectangular strips.  Two dimensional geometries have only one length scale ($L_x = L_y$), while   pseudo-one-dimensional geometries have one extra length scale (both  $L_x$ and $L_y$).  This complicates the phase diagram, producing an insulating phase along the weak-disorder edge of the conductance plateau.  This phase is only one of many finite size effects which blur and distort the phase diagram.   In a strip with short dimension $L_x$ and long dimension $L_y$, the finite size effects are determined by $L_x$.  The maximum value of $L_x$ until now has been $L_x=300a$, in a  paper which used $L_x=130a$ data to determine the TAI phase diagram  (Ref. \onlinecite{Chen11}).  Other studies have considered $L_x=12a$ (Ref. \onlinecite{Zhang11b}), $L_x=64a$ (Ref. \onlinecite{Yamakage11}), $L_x=50a$ (Ref. \onlinecite{Prodan11}), and $L_x = 150a$ (Ref. \onlinecite{Li09}).  While these lengths are sufficient to measure scattering physics, in much of the phase diagram they are unable to probe the localization length scale.  Moreover  finite size effects are quite large, blur the phase boundaries, and prevent a precise understanding of individual phases.

  In section \ref{FiniteSizes} of this  paper we obtain a clear picture of the physics at length scales $L_x = L_y   \leq 280 a$.  This brings the scattering length scale into sharp focus, and allows us to see clearly both a scaling region where the conductance plateau is accompanied by a bulk gap and a larger  scaling region where the plateau persists without an accompanying bulk gap.   (We use the term "scaling region" rather than "phase" only because  this black and white gapped vs. ungapped dichotomy must be refined a bit in larger systems.) We use the two-dimensional $L_x = L_y$ geometry, and we change the system size by a factor of four ($L = 70, 140, 280a$) to get a clear picture of finite size effects.   The central result of this section is the TAI phase diagram, Figure ~\ref{fig:PhaseDiagram1}.  We also discuss the eigenstates in the conductance plateau: both their peculiar property of growing rather than shrinking when the disorder strength is increased, and their fractality.
  
In section \ref{InfiniteVolumeLimit} we turn our focus toward the localization length scale and increase the system size to  $L = 1120 a$.   The phase boundaries become more clearly defined, and allow us to rule out the existence of extra phases at the edges of the TAI conductance plateau.  At the same time the phase boundaries move because localization effects are becoming important.   We also find that the bulk  band gap is invaded by localized bulk states.  The density of bulk states in the gap is exponentially small as is typical for Lifshitz tails, which allows the gapped and ungapped scaling regions to be distinguished from each other.  Lastly we report the  conductance distributions on various phase boundaries and compare them with known critical distributions for the Integer Quantum Hall and Quantum Spin Hall systems.

\subsection{The TAI model\label{TAIModel}}
Following the original TAI paper \cite{Li09}, we study the BHZ tight binding model   \cite{Bernevig06a} of 2-D topological insulators.  The mass can be tuned by changing the sample thickness, and we choose a positive mass $M = 1$ meV so that there is no band inversion and no edge states at zero disorder.  The BHZ model's momentum representation is:
\begin{eqnarray}
\mathcal{H} & = & \begin{bmatrix} h(\vec{k}) & 0 \\ 0 & h^*(-\vec{k}) \end{bmatrix}, \; \; D   =  512 meV \, nm^2, \; B= 686 \, meV \, nm^2, \; A = 364.5 meV \, nm, \; a = 5 \, nm, \; M = 1 \, meV,
\nonumber \\ h(\vec{k}) & = &   (A/a)  \sigma_x \sin (k_x a) +    (A/a) \sigma_y \sin (k_y a) +  M \sigma_z  + 2 (\sigma_z B/a^2 + I \, D/a^2) (2 - \cos(k_x a) - \cos(k_y a))
\end{eqnarray}
The basis has four orbitals: the first two are $s$ and $p$ orbitals with spin $s_z = 1/2$, and the last two are $s$ and $p$ orbitals with spin $s_z = -1/2$. $\sigma_{x,y,z}$ are the Pauli matrices. The Hamiltonian conserves the $s_z$ component of the spin, which is protected by the combination of inversion symmetry and  axial symmetry around the growth axis.  This symmetry can be broken by a quantum well or gate electrode \cite{Yamakage11}, and also can be broken by disorder.  We follow the original TAI papers \cite{Li09, Jiang09, Groth09}  which preserved the $s_z$ symmetry and  introduced nonmagnetic on-site disorder randomly distributed in the interval $\left[ -W/2, \, W/2 \right] $, where $W$ is the disorder strength.   With this disorder the BHZ model factorizes into two independent $2 \times 2$ Hamiltonians which are time-reversal counterparts of each other.  Therefore all of our   calculations of the conductance, eigenvectors, and eigenvalues concern themselves only with the $s_z = 1/2$ sector, which is  governed by a $ 2 \times 2$ Hamiltonian.  The conductances and  densities of states  which we report are exactly $1/2$ of the correct values that are obtained when all four orbitals are included.  

It is very important to recognize that  the BHZ model's $s_z$ symmetry substantially changes its localization physics \cite{Groth09, Prodan11, Onoda07, Obuse07}.   If kinetic or disorder terms which break the $s_z$ symmetry are added to the BHZ model, then factorization is disallowed and one must use the original $4 \times 4$ model which belongs to the symplectic symmetry class and can exhibit metallic phases \cite{Yamakage11}.  In contrast, the factorized $2 \times 2$ model belongs to the unitary symmetry class, the same as systems exhibiting the integer quantum hall effect.  Unlike symplectic models, unitary models are always localized in infinite 2-D systems.  

The principal focus of this paper is on the TAI phase diagram in a two-dimensional $L_x = L_y$ geometry.  This question is most clear in an isolated sample without any leads, where it can be studied by analyzing  eigenstates and eigenvalues.   With the exception of our conductance data, all of our numerical results come from the leads-free geometry.  However for the conductance we use  leads with width $w = L_x = L_y$ adjoining  two opposite sides of the the disordered sample.  As discussed in section \ref{DelocalizationAndFractality}, the introduction of leads does not change the phase diagram significantly. We evaluate the conductance using the Caroli formula\cite{Caroli71,Meir92} $G = -\frac{e^2}{h}{Tr}((\Sigma^r_{L}-\Sigma^a_{L}) G^r_{LR} (\Sigma^r_{R}-\Sigma^a_{R}) G^a_{RL})$, where $G^a, G^r$ are the advanced and retarded Green's functions connecting the left and right leads and $\Sigma_{L,R}$ are the self-energies of the leads. We evaluate the lead self-energies using the well-known iterative technique developed by Lopez Sancho et al \cite{Lopez84}.    Following References \onlinecite{Jiang09} and \onlinecite {Groth09}, we use semi-infinite leads that are described by the BHZ Hamiltonian  and are free from disorder. Because the mass  $M=1$ meV is positive, in the leads the bulk band gap does not contain any edge states.  Therefore all conducting channels in the leads are bulk channels, and the leads do not conduct when the Fermi energy is in the gap $|E_F| < M$.  Errors associated with the leads will be discussed in Appendix \ref{NumericalErrors}.

\section{The Phase Diagram in Systems of Size $L \leq 280 a$\label{FiniteSizes} }
In this section we will defer the question of the infinite  size limit and focus instead on the physics and phases in finite systems with $L \leq 280 a$.  First we will show that the TAI conductance plateau contains both a gapped and an ungapped scaling region,  and  will show that the CPA gives a good description of the gapped scaling region but not the ungapped scaling region.  The centerpiece of this section is the phase diagram, which shows all the phases and their boundaries.  Lastly we will study the spatial structure of the bulk states.

Because our Hamiltonian belongs to the unitary ensemble, we expect that in the infinite volume limit all bulk states are localized and non-conducting.    Nonetheless  for systems of size $L \leq 280 a$ a large portion of the phase diagram exhibits an average conductance that is proportional to the system size $L$ for all $L \leq 280 a$, implying that the bulk states are extended and conducting even at $L = 280 a$.   We will call this behavior "bulk conduction."

   \subsection{The TAI conductance plateau}
\begin{figure}
\begin{center}
  \includegraphics[width=16cm]{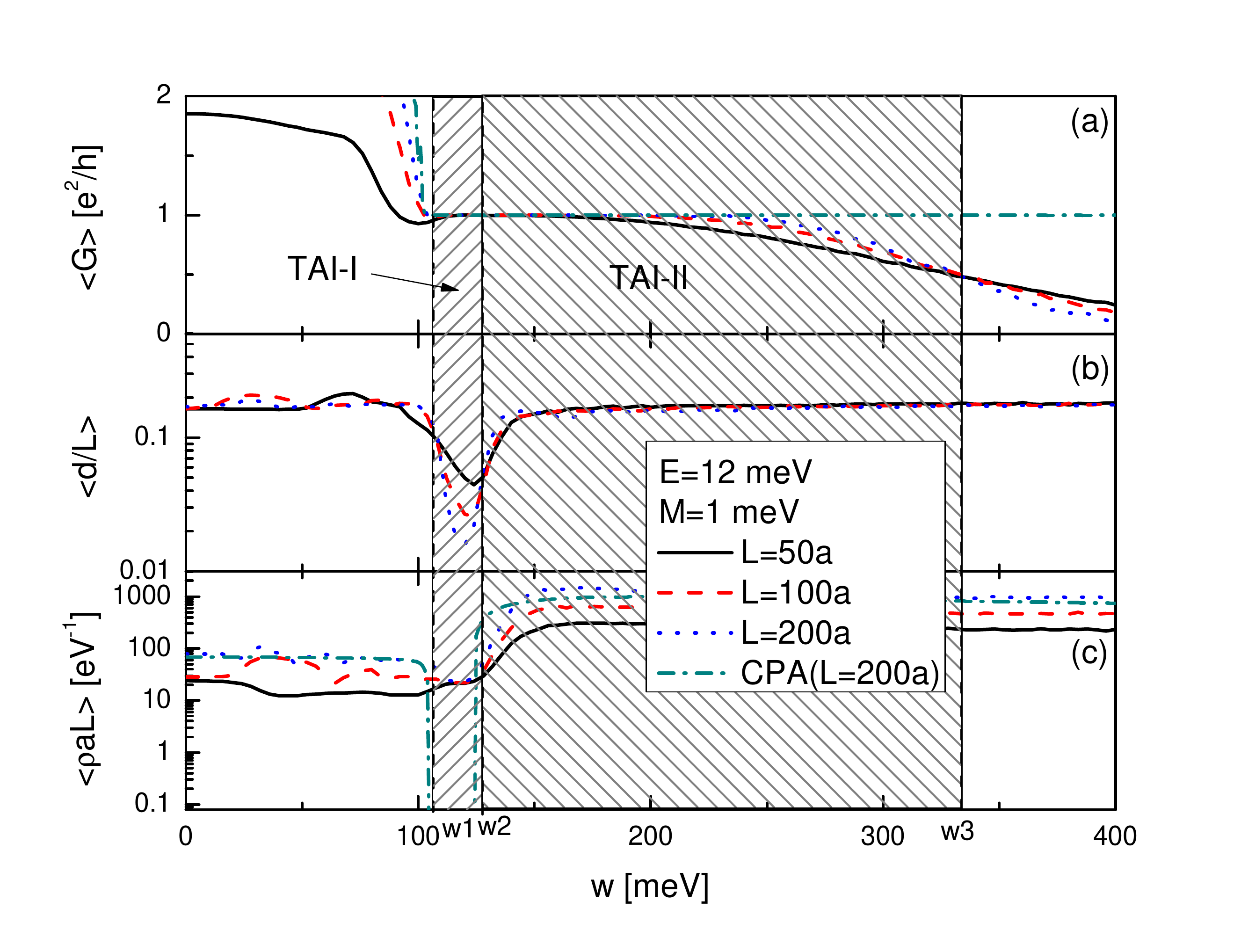}
   \end{center}
   \caption{Two TAI scaling regions: TAI-I with only edge states, and TAI-II with both edge states and localized conducting states. Lines $w1$ and $w3$ mark the edges of the TAI conductance plateau, and $w2$ marks the boundary between the gapped TAI-I scaling region on the left and the ungapped TAI-II scaling region on the right.  Panes a, b, and c show respectively the conductance, the scaled distance from the edge, and $L \,a$ times the local density of states. The CPA (L = 200a) lines report the results of the coherent potential approximation.  The Fermi energy is fixed at $E = 12$ meV. }
   \label{fig:E12}
\end{figure}

Figure ~\ref{fig:E12}a demonstrates TAI conductance quantization, which is manifested as a plateau in the average conductance $\langle G \rangle = 1$ with boundaries at the disorder strengths $w1 \approx 105$ meV and $w3 \approx 334$ meV.  The Fermi energy is kept fixed at $E = 12$ meV in this figure. The large-disorder edge at $w3$ is a transition from quantized conductance to the localized phase.   As is typical of second-order phase transitions,  it is broadened at finite system sizes, narrowing the TAI conductance plateau.  When the system size is increased from $L = 50 a$ to $L=200 a$ the transition narrows, and therefore $\langle G \rangle$ increases towards one on the left of $w3$ and decreases towards zero on the right.  This produces a  crossing point  that allows us to estimate the position of the true phase transition in an infinite system.   In this limit the conductance transition at $w3$ is  a step function which changes directly from one  to zero. In other words, at $w3$  the conducting edge states are abruptly destroyed in a process analogous to the integer quantum hall effect.   

The small disorder edge at $w1$ is a transition from bulk conduction to quantized conductance. Here the conductance exhibits a meeting point, and the rise to the left of that meeting point   becomes more and more abrupt as the system size is increased.  The trend is toward a very sharp conductance transition at $w1$ that moves directly from  bulk conduction to $ \langle G \rangle = 1$.

     In Figure ~\ref{fig:E12}b we calculate eigenstates, and show their scaled average distance \footnote{We calculate the distance of $\vec{x}$ from each of the four edges.  $d(\vec{x})$ is the minimum of these four distances.} from the edge, $\langle d/L \rangle =  L^{-1} \int {d\vec{x}} |\psi(\vec{x})|^2 d(\vec{x})$.  Between $ \approx 108$ meV and $w2 \approx 129$ meV  this observable converges toward zero as the system size is increased from $L=50a$ to $L=200a$, indicating that the eigenstates are edge states.  However outside of this range  $\langle d/L \rangle$ is independent of the system size, indicating that the eigenstates are bulk eigenstates, distributed in the interior of the sample.  

We conclude that the TAI conductance plateau is composed of two scaling regions, one with a bulk band gap, and the other without.    In the  TAI-I region to the left of $w2$ the bulk gap leaves only edge states, which are responsible for conduction.  However there is a second TAI-II region to the right of $w2$ with both bulk states and edge states; it has no bulk gap.  Since the number of bulk states scales with $L^2$ while the number of edge states scales with $L$, the bulk states dominate $\langle d/L \rangle$. The bulk states are localized, maintaining the TAI  conductance quantization $\langle G \rangle = 1$.  Figure ~\ref{fig:E12}c's plot of the average local density of states $\langle \rho \rangle$ confirms this picture: in the TAI-I (gapped) region $\langle \rho L \rangle$ is independent of the system size $L$, as is expected of edge states.  In the TAI-II (ungapped) region $\langle \rho L \rangle$ is proportional to $L$, indicating that bulk states are far more numerous than edge states.  

\begin{figure}
\begin{center}
  \includegraphics[width=16cm]{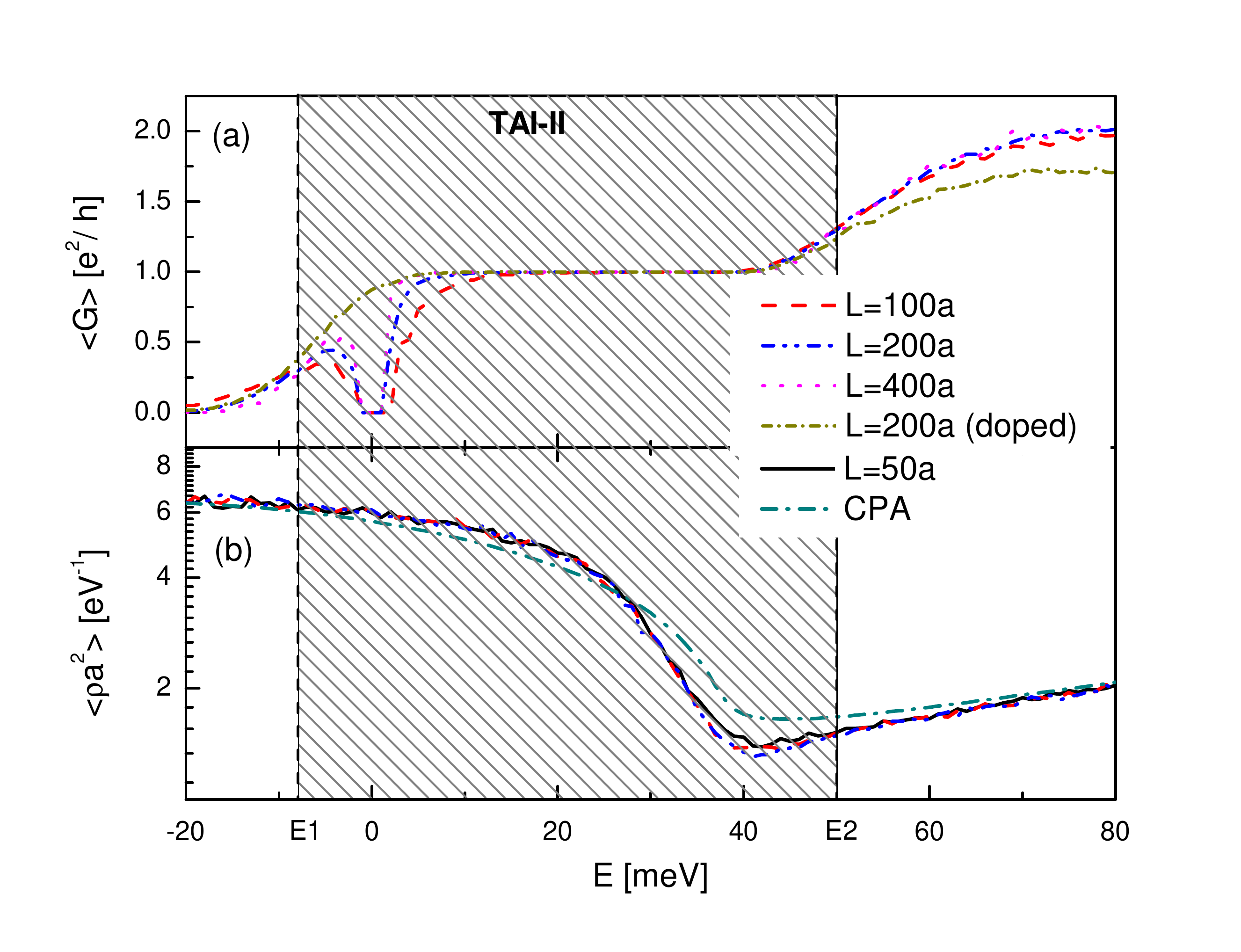}
   \end{center}
   \caption{TAI Conductance Quantization without a Band Gap. The disorder strength is fixed at $W = 200$ meV.  $E1$ and $E2$ label crossing points which mark the edges of the conductance plateau.  The local density of states $\rho$ is independent of system size $L$, indicating that it is dominated by bulk states. The conductance gap near $E=0$ is caused by the disorder-free leads, which do not conduct at $|E| \leq 1$ meV.   The "L=200a(doped)" line shows clearly that when the leads are doped to $E= 25 $ meV the conductance plateau extends to the $E1$ crossing point.  } 
   \label{fig:w200}
\end{figure}

A bulk band gap is not necessary for TAI conductance quantization.  Figure ~\ref{fig:w200}a displays the conductance at a fairly large value of disorder, $W = 200$ meV.  The boundaries of the TAI conductance plateau coincide with the  clear crossing points at $E1 \approx -8$ meV and $E2 \approx 50$ meV.   Figure ~\ref{fig:w200}b shows that  the  local density of states $\rho$ is independent of the system size, implying that there is no bulk band gap.   The entire TAI conductance plateau lies in the TAI-II scaling region, which  is characterized by the coexistence of conducting edge states with localized bulk states.  This shows that a mobility gap - even in the absence of a band gap - is sufficient   for producing TAI conductance quantization.

Figure ~\ref{fig:w200} crosses the weak-disorder boundary of the TAI conductance plateau at line $E2$, but this crossing is not accompanied by a bulk band edge.    This establishes that the weak-disorder boundary is accompanied by a bulk band edge only in the gapped TAI-I region, not in the ungapped TAI-II region.  In the TAI-II region the bulk states exhibit an Anderson transition as the weak-disorder boundary is crossed, and  localization is responsible for preventing bulk conduction.   In section \ref{FateOfTheGap} we will show that  in the TAI-I gap there is  an exponentially small density of localized bulk states; the TAI-I weak-disorder boundary is marked by an Anderson transition from bulk conduction to  localized bulk states.   Zhang et al \cite{Zhang11b}  found these localized bulk states in very small $8 a \times 8 a$ systems.  Their localization is responsible for the  TAI conductance plateau where only edge states conduct.   In section \ref{TransitionToPlateau} we will study systems as large as $1120 a  \times 1120 a$ and find that the weak-disorder boundary of the conductance plateau is always accompanied by a mobility edge, both in the gapped TAI-I region and in the  TAI-II region where there is no gap and no bulk band edge.

The gapped TAI-I scaling region and its boundary with the ungapped TAI-II scaling region can be understood entirely within the  coherent potential approximation \cite{Soven67, Taylor67}.   Figure ~\ref{fig:E12}c and Figure ~\ref{fig:w200}b include the CPA density of states $\rho$, which provides excellent predictions of the true density of states everywhere except in the gapped TAI-I region.      The CPA  is a mean field approximation that does not include edge physics, so inside the TAI-I region it predicts only the gap $\rho = 0$ and not the edge states.  The CPA density of states nonetheless makes a  spectacularly good prediction of all boundaries of the TAI-I region, including its boundary with the neighboring TAI-II region.

At  an intermediate step in calculating the CPA density of states one obtains the self-consistent mean-field Hamiltonian $H_{CPA}$, whose real part can be interpreted as a renormalized band gap and renormalized Fermi level.  The Fermi level's position (inside or outside the gap) can predict the weak-disorder boundary of the TAI conductance plateau \cite{Groth09}.  This predictor duplicates the CPA density of states' success with the gapped TAI-I region's weak-disorder boundaries, but unlike the DOS it does not find the  boundary with the TAI-II region.

Our picture of two scaling regions distinguished by bulk vs. edge scaling is based on analysis of data at finite system sizes $L = 50, \, 100, \, 200a$.  It matches very well with the CPA, which incorporates only physics at the scattering length scale, and neglects physics at the much longer  length scale of localization.   We conclude that the gapped TAI-I  region and its boundary with the ungapped TAI-II region is caused entirely by  short-distance scattering physics.     Our picture of the two regions will require some refinement  when the system size is increased,  as will be discussed in section  \ref{InfiniteVolumeLimit}, but scattering physics and  the bulk band gap will still strongly influence the phase diagram.

\subsubsection{The Coherent Potential Approximation Hamiltonian}

   We have re-used the self-consistent mean-field Hamiltonian $H_{CPA}$ in a recursive Green's function calculation of the conductance, omitting the imaginary part of $H_{CPA}$ which is anti-Hermitian and which destroys the CPA's predictive power at almost all values of the disorder strength $W$.  The CPA conductance, calculated while omitting the imaginary part of $H_{CPA}$, is shown in Figure ~\ref{fig:E12}a.    It is correct at disorder strengths smaller than $w3$, including the TAI-I region where there are only edge states.  However, the CPA conductance fails to predict the localization of the edge states at $w3$, and remains quantized even at very large disorder.  In addition, the CPA Hamiltonian $H_{CPA}$ does not predict the localized bulk states seen in the ungapped TAI-II region and does not distinguish the TAI-I and TAI-II regions; it predicts that both the density of states $\rho$ and $\langle d / L \rangle$  retain edge values at all disorders larger than $w1$.    In summary, the mean-field Hamiltonian $H_{CPA}$  correctly predicts observables both in the bulk conducting phase and in the  gapped TAI-I region, but fails to predict the localized bulk states seen in  the  TAI-II region and in  the ungapped insulating phase, and also fails to find the large-disorder edge of the TAI conductance plateau.

\subsection{Phase Diagram}

\begin{figure}
\begin{center}
  \includegraphics[width=16cm]{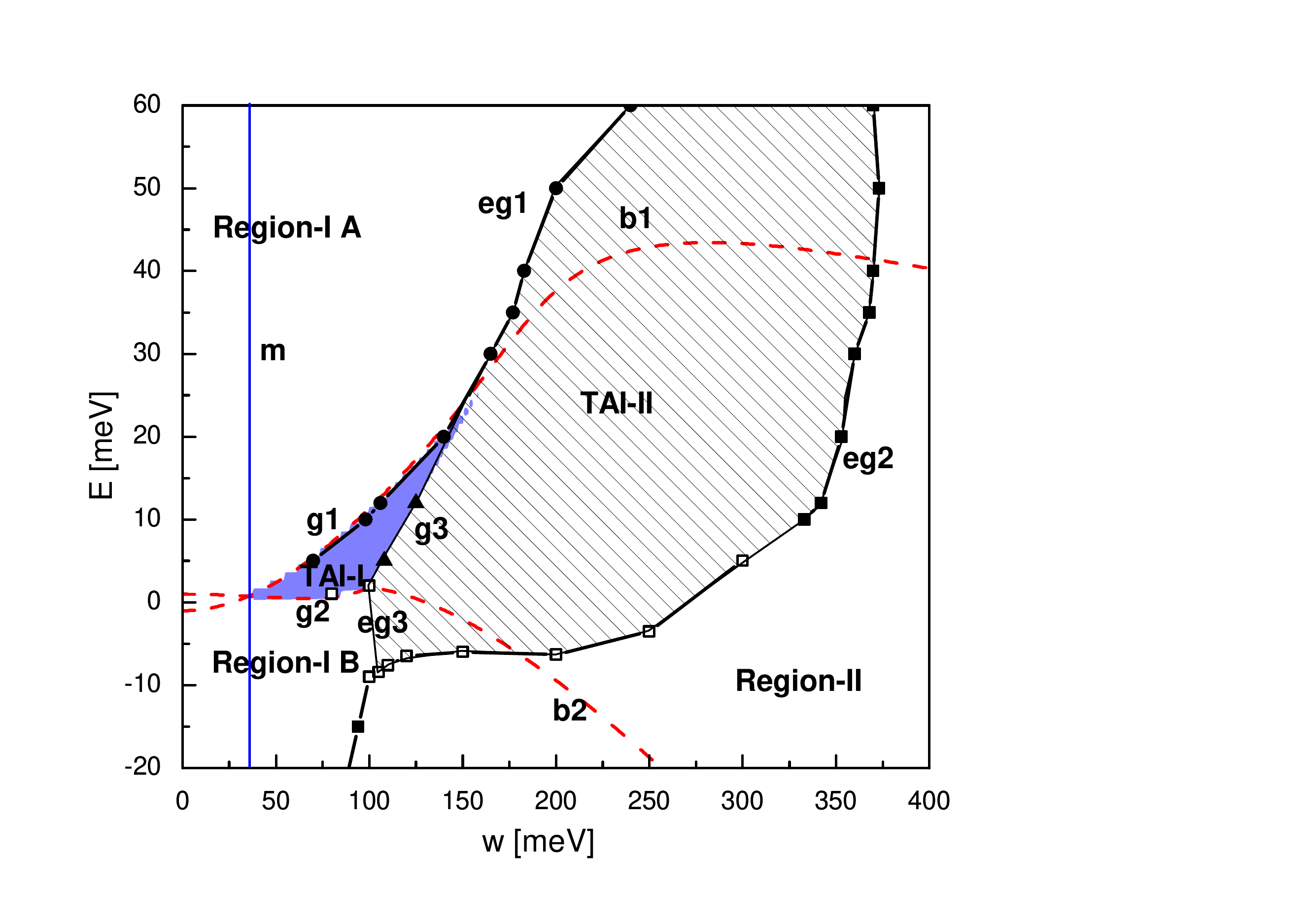}
   \end{center}
   \caption{The TAI phase diagram at mass $M=1$ meV.  In Regions I A and I B the bulk states conduct.  An Anderson transition separates Region I B from Region II, where the system is insulating.  The TAI conductance plateau includes both the gapped TAI-I region (colored blue) and the ungapped TAI-II region (hatched lines.)  The region inside the dashed lines to the left of line $m$ is gapped and has no surface states.  This phase diagram reflects only physics at length scales $L \leq 280 a$.  }     \label{fig:PhaseDiagram1}
\end{figure}

The phase diagram is reported in Figure ~\ref{fig:PhaseDiagram1}.   Each data point represents a crossing point or meeting point which was determined by comparing observables  at three different system sizes as described in our discussion of  Figures  ~\ref{fig:E12} and  ~\ref{fig:w200}.   Numerical errors will be discussed in Appendix \ref{NumericalErrors}.  The triangles along line $g3$ were obtained from crossing points of the scaled distance from the edge $\langle d / L \rangle$, and all other data points were obtained from  the average conductance $\langle G \rangle$. The open boxes report crossing points observed on vertical lines keeping $W$ fixed, while the filled data points were determined on  horizontal lines keeping $E$ fixed.  

We have connected the data points with lines. Lines  $eg1, \, g1$, $g2$, and $eg3$ represent a bulk conduction-to-quantized transition which separates the TAI plateau from Region-I A and Region-I B.  In Region-I A  the conductance scales with system size $L$, signifying bulk conduction.  Region-I B also exhibits bulk conduction, and is separated from the insulating (zero conductance) Region-II by an Anderson transition.  Line $eg2$  represents a quantized-to-insulator transition, and separates the localized phase in Region-II from the conductance plateau.  Contrary to Chen et al \cite{Chen11}, we see no evidence of an intervening metallic phase along $eg2$; we will discuss this further in section  \ref{LocalizationTransitions}.  We have not done a careful analysis of the phase diagram above $E = 60 $ meV, but we have found that at  $E = 100$ meV there is no conductance plateau.  This suggests that lines $eg1$ and $eg2$ meet and merge into an Anderson transition somewhere between $E = 60 $ meV and $E = 100 $ meV. Line $g3$ separates the gapped TAI-I region from the ungapped TAI-II region, which is marked by hatched lines.

Our phase diagram includes also the CPA predictions for the phase structure.  Both the area with blue shading and the small area between dashed at lines at $W < 35$ meV are characterized by a bulk band gap, as calculated \footnote{We determined the boundaries of the shaded region by comparing the CPA Local DOS to $10^{-8}$.} using the CPA.  The CPA bulk band gap succeeds brilliantly in predicting the boundaries of the gapped  TAI-I region (shaded blue): its predictions agree with the data points on lines $g1, \, g2$, and $g3$ to within $2$ meV.  This allows us to estimate the positions of two tricritical points at the ends of line $g3$: they lie at $ (W \approx 101, E \approx 2)$ meV and  at $(W \approx 158, E \approx 25)$ meV.

We mark the gap closure and rebirth at $W \approx 35$ meV with line m.  To the left of line m  the material is not in a topological phase, because the mass $M = 1$ meV is positive.  To the right of line m a bulk band inversion begins and  a topologically protected band of edge states appears and persists until  $W \approx 370$ meV.  At $W \approx 370$ meV  the edge band collapses, the bulk band inversion terminates, and the material is no longer a topological insulator.  

Line $eg3$ separating the bulk conducting phase in Region-I B from the quantized TAI-II region  is merely a guide to the eye; our data do not allow us to decide exactly where the bulk conducting phase terminates.     Along the $W = 100 $ meV line we see an interval with bulk conduction sandwiched between the quantized TAI-I region and the insulating phase, while at $W = 105,110 $ meV this evidence  disappears quickly. This small portion of the phase diagram is extremely complex, with four phases  and regions and two tricritical points near each other.  Analysis is complicated even further by the leads, which if left undoped display very large finite size effects near the $E=0 $ meV axis.  

Line $eg2$ marks the lower boundary of the band of edge states.   It seems unlikely that this boundary follows the edge of the conductance plateau along $eg3$ and $g2$, since this would involve a discontinuous transition between the $eg2$ line which runs along $E \approx -8 $ meV and $g2$ line at $E = 2 $ meV.   More likely the band of edge states extends into the bulk conducting Region-I B but its contribution to the conductance is hidden by the bulk conduction signal.    Similarly, the upper boundary of the edge band may extend into Region-I A, above lines $g1$ and $eg1$.  We conjecture that  the lower boundary of the edge band may interpolate smoothly between line $eg2$ and the point  where the bulk band gap closes and reopens.

$b1$ and $b2$ mark lines where the Fermi level passes through the edges of the bulk band gap - using a Fermi level and band gap that have been derived from the self-consistent  mean-field CPA Hamiltonian.    These predictors fail to predict the TAI-II portion of the weak-disorder boundary of the TAI plateau.  For instance, line $b1$ predicts that the $W = 200 $ meV line exits the TAI conductance plateau at $E = 38 $ meV.  Referring to figure ~\ref{fig:w200}, we find that although there is minimum in the local density of states at $E = 41$ meV, the true phase boundary (a crossing point in the conductivity) is located at $E = 50 $ meV.   Lines $b1$ and $b2$  do not seem to be good predictors of the ungapped TAI-II region's boundaries.

\subsection{Bulk State Delocalization and Fractality\label{DelocalizationAndFractality}}

\begin{figure}
\begin{center}
  \includegraphics[width=16cm]{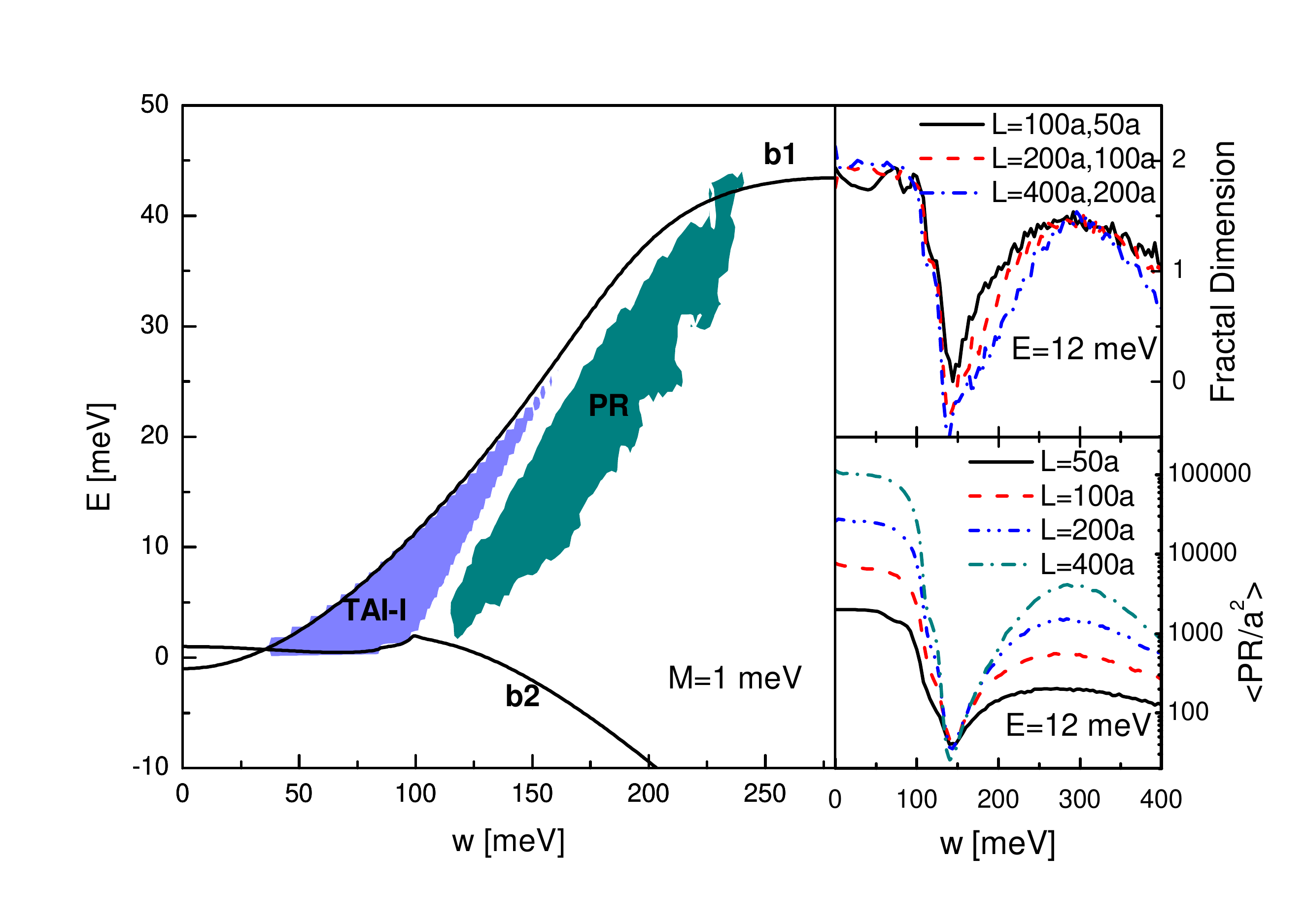}
   \end{center}
   \caption{Eigenstate hyperlocalization near the mean-field band edge.  Inside the colored "PR" area the average PR is very small.  (In this plot we show the region where $\langle PR \rangle \leq e^5 a^2 \approx 148 a^2$ at system size $L= 200a$.)  If we omit  surface states from the average, then we find that the bulk states are hyperlocalized in both the gapped TAI-I region and the PR region.    Lines $b1$ and $b2$ show where the CPA-adjusted Fermi level crosses the CPA-adjusted band edges, and match well with the bulk hyperlocalized region.   The lower right inset shows the average participation ratio at four system sizes, while the upper right inset shows the fractal dimension.  The Fermi energy is fixed at $E_F = 12 $ meV in both insets. }
   \label{fig:Hyperlocalized}
\end{figure}

Remarkably, eigenstates grow when the disorder strength is increased throughout a large portion of the ungapped TAI-II region.   This is the reverse of most systems: usually  when the disorder strength is increased eigenstates  shrink and move toward localization.    We calculated eigenstates of the TAI hamiltonian and measured the participation ratio (PR) $PR^{-1} = \int d^2\vec{x} |\psi(x)|^4$, which is a measure of the eigenstate volume.  The PR ranges from one for a fully localized eigenstate to the system volume $V$  for a fully extended eigenstate.   The  lower right inset of Figure ~\ref{fig:Hyperlocalized}  reports the average PR, which includes contributions from both bulk and edge states.   Near the transition from bulk to quantized conductance the average PR drops precipitously to very small values around $ 40 a^2$, indicating that the  eigenstates become very localized.  This region of hyperlocalization is very small: it adjoins a broad peak that rises through the TAI-II region, reaches a maximum  near the quantized-insulator transition, and falls in the insulating phase.  In other words the eigenstates steadily grow with increasing disorder, reach their maximum size near the quantized-insulator transition, and then begin to localize again. The width of the peak decreases as the system size is increased.

Lines $b1$ and $b2$ show  where the mean-field Fermi level passes through the edges of the mean-field bulk band gap. Figure ~\ref{fig:Hyperlocalized} compares these lines to a contour of the average PR at $\langle PR \rangle = e^5 a^2 \approx 148 a^2$.  Earlier we found that the transition from bulk to quantized conduction is  predicted by $b1$ and $b2$ only near the gapped TAI-I region; $b1$ and $b2$   fail  at the edge of the ungapped TAI-II region.    Now we find that the hyperlocalized region (labeled PR) is predicted well by line $b1$ even in the TAI-II region.     In Figure ~\ref{fig:Hyperlocalized} we include both edge and bulk states in $\langle PR \rangle$, so in the gapped TAI-I region we see values typical of delocalized edge states, much larger than $148 a^2$.  We have calculated the average PR of only the bulk states, omitting edge states, and have found that the bulk states are hyperlocalized  throughout the entire TAI-I region, with a  minimum  $\langle PR \rangle$ of about $10 a^2$.     In summary, lines $b1$ and $b2$ seem to be good predictors of the bulk hyperlocalized region but not of the phase transition to the TAI-II region.

The upper right inset of Figure ~\ref{fig:Hyperlocalized}  shows the fractal dimension $d_2$, which is the derivative of $\ln \langle PR \rangle$ with respect to the logarithm of the system size $L$; we report $d_2(400a,200a) = ( \ln \langle PR (L=400a) \rangle -\ln \langle PR (L=200a) \rangle) /\ln 2$, $d_2(200a,100a)$, and  $d_2(100a, 50a)$.  Non-integer values indicate fractal states.  Regarding non-fractal states,   the integers $d_2 = 2, \; 1$, and $0$ respectively indicate conducting extended bulk states, edge states, and localized states.  Figure ~\ref{fig:Hyperlocalized} shows that throughout the TAI-II region the bulk eigenstates are fractals.  Since their size is a power of the system volume, they are sensitive to the boundary conditions.  It is remarkable that  the average conductance $\langle G \rangle $ remains quantized, indicating that these bulk states do not conduct even though they are edge-sensitive. 

 Like the   participation ration $\langle PR \rangle$, the fractal exponent $d_2$ forms a broad peak centered near the quantized-localized phase transition, and the peak width decreases as the system size is increased.   Along the $E = 12 $ meV line we find that the peaks in both $d_2$ and $\langle PR \rangle$ are located near $W = 290 $ meV and don't move much as the system size is increased.  This disagrees with our conductance data, which indicates that the quantized-insulator phase transition is located at $W > 300 $ meV.  This discrepancy could be evidence for a systematic disagreement between data based on the conductance (obtained with infinite leads)  and data based on diagonalization of the Hamiltonian (obtained with no leads).  However it seems more likely that the discrepancy is a finite size effect: we will see in section \ref{LocalizationTransitions} that the conductance crossing point shifts from $W \approx 380 $ meV to $W \approx 320$ meV as the system size is increased from $L = 50 a$ to $L = 400a$.  Most likely the peaks in both $d_2$ and $\langle PR \rangle$ coincide with the quantized-localized phase transition.
 
Groth et al \cite{Groth09} measured the scaling exponent $\nu = 2.66 \pm 0.15$ at the quantized-insulator phase transition,  which is compatible with the known value\cite{Slevin09} of $2.59$ for the integer quantum hall effect; they suggested that the TAI quantized-insulator transition belongs the IQH universality class.  Here we report that the  maximum value of the fractal exponent $max(d_2)$  near the quantized-insulator phase transition is approximately $d_2\approx1.5$, and is consistent with the value of $1.5 \pm 0.1$ seen in previous studies of the fractal exponent at the integer quantum hall transition \cite{Huckestein94,Terao96,Klesse97,Huckestein99}.

\section{Infinite Volume Limit\label{InfiniteVolumeLimit}}
In the previous section we confined ourselves to system sizes $L \leq 280 a$, and obtained a phase diagram appropriate to that length scale. This brought the scattering physics into sharp focus, but finite size effects blurred and shifted the phase boundaries. In disordered systems the physics at the localization length scale can be quite different from the physics at smaller length scales.  In 2-D systems the localization length scale can be extremely large.  Therefore it is very important to attempt some extrapolation of the phase diagram to larger system sizes. 

Already the $280 a \times  280 a$ systems of the previous section compare well to previous TAI studies, but in the present section we calculate certain observables in systems that are four times bigger, multiplying the computational time by $256$.  We were able to do this for several reasons. Firstly, fluctuations in $G$ drop to zero at the transition from bulk conduction into the TAI conductance plateau, so in this region we need only very small data sets.  Secondly, we concentrate our efforts on a four specific points in the phase diagram, with each point lying on a phase boundary.  Lastly, we content ourselves with visual examination of crossing points instead of systematic fitting which would have required high statistics.  The payoff for these choices is clear information about the evolution of the phase diagram at large system sizes.

\subsection{Transition from Bulk conduction to the TAI Plateau\label{TransitionToPlateau}}
\begin{figure}
\begin{center}
  \includegraphics[width=16cm]{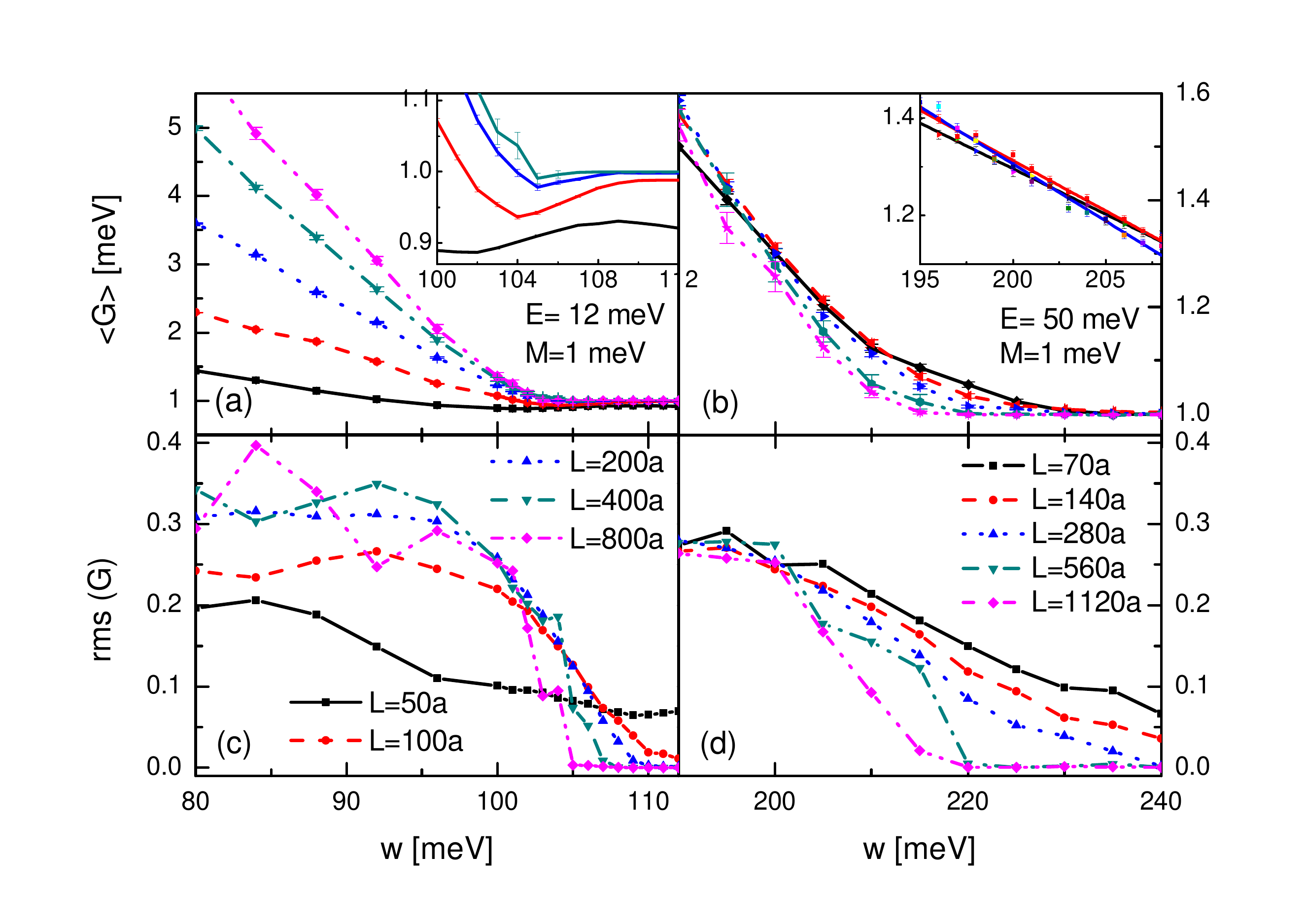}
   \end{center}
   \caption{Phase transitions from the bulk conducting phase to the TAI-I and TAI-II scaling regions. Panes a and c show the transition to the gapped TAI-I scaling region, while panes b and d show the transition to the ungapped TAI-II scaling region.  The insets show closeups of the conductance data.  The straight lines in the upper right inset are least squares linear fits to the data.  }
   \label{fig:PhaseTransitionCloseups}
\end{figure}

Figure ~\ref{fig:PhaseTransitionCloseups} examines the boundary between the bulk conducting phase and the TAI conductance plateau, including  both the gapped TAI-I  scaling region and the ungapped TAI-II  scaling region.   Figure  ~\ref{fig:PhaseTransitionCloseups}b shows the transition from bulk conduction to the ungapped TAI-II region.  At this transition all of the bulk states become localized.  There is a clear crossing point near $W \approx 200 $ meV, indicating scale-invariant physics typical of disorder-induced phase transitions.   

Unlike the TAI-II transition,  the transition to the  gapped TAI-I region (Figure   ~\ref{fig:PhaseTransitionCloseups}a) is a meeting point.  Moreover, our data at $L = 50, 100, 200, 400$, and $800 a$ shows very clearly that  the meeting point is converging to $W \approx 106 $ meV.  The conductance dip seen so clearly at $L= 50a$ shrinks and moves to the right; it is a finite size effect.  The meeting point and its excellent convergence indicate that the  gapped TAI-I physics has a finite length scale, and therefore point toward physics more complex than a simple bulk band edge.    Bulk states inside a band gap have a decay length.  This decay length is a function of the distance  $E_F - E_{band}$ of the Fermi energy from the band edge, and diverges at the band edge where $E_F - E_{band} = 0$.  Our data indicates that there is no such diverging length scale at the weak-disorder edge of the gapped TAI-I region.   

Figures   ~\ref{fig:PhaseTransitionCloseups}c and d show the second moment of the conductance at the transitions from bulk conductance to the TAI-I and TAI-II regions.   In the bulk conducting phase the second moment is finite and increases with the system size.  Both graphs show that  $rms(G)$ converges to zero in the TAI conductance plateau.   As the system size is increased the transition in $rms(G)$ sharpens and moves toward coinciding with the transition in $\langle G \rangle$.  This corroborates our determination of the bulk-to-quantized phase transition, and also allows us to measure $\langle G \rangle$ precisely with small statistics.

Because our Hamiltonian is a member of the unitary symmetry class, we expect that the bulk Anderson transition - the extinction of bulk conduction - should move toward zero disorder $W = 0$ as the system size is increased.  Therefore we should see the conductance curves shift toward smaller disorder as the system size is increased.    Moreover, in large enough systems the bulk Anderson transition should  detach from the  TAI conductance plateau, and one should see a "notch" of zero conductance grow and  separate the bulk conduction phase from the TAI conductance plateau.   Figure ~\ref{fig:PhaseTransitionCloseups}a does not agree with these expectations - there is no evidence of any movement or notch at the edge of the gapped TAI-I region.   The inset of Figure ~\ref{fig:PhaseTransitionCloseups}b indicates that the edge of the TAI-II region is moving, but the movement is quite small.  There is no sign of a conductance notch anywhere, despite searches at system sizes at large as $L= 1120 a$.  We find no evidence that the weak-disorder boundary of the TAI conductance plateau can exist independently of a mobility edge.  In summary, the transition from bulk conductance to the TAI conductance plateau is very stable under increases of the system size.   

  This stability is peculiar to the $2-D$ square geometry which we study.  With the exception of Ref. \onlinecite{Groth09}, all other studies of the TAI  conductance  have invariably calculated conduction along quasilinear strips which are much longer than their width.   In this geometry bulk conduction is controlled by the ratio of the strip length to the localization length, while edge conduction is controlled by the ratio of the strip width to the localization length.  As a result a $\langle G \rangle = 0$ conductance notch between the bulk and quantized phases is easy to reproduce.  In a true $2-D$ geometry the two ratios are identical, resulting in the notch-less physics which we observe.

\subsection{Localization Transitions\label{LocalizationTransitions}}
  
\begin{figure}
\begin{center}
  \includegraphics[width=16cm]{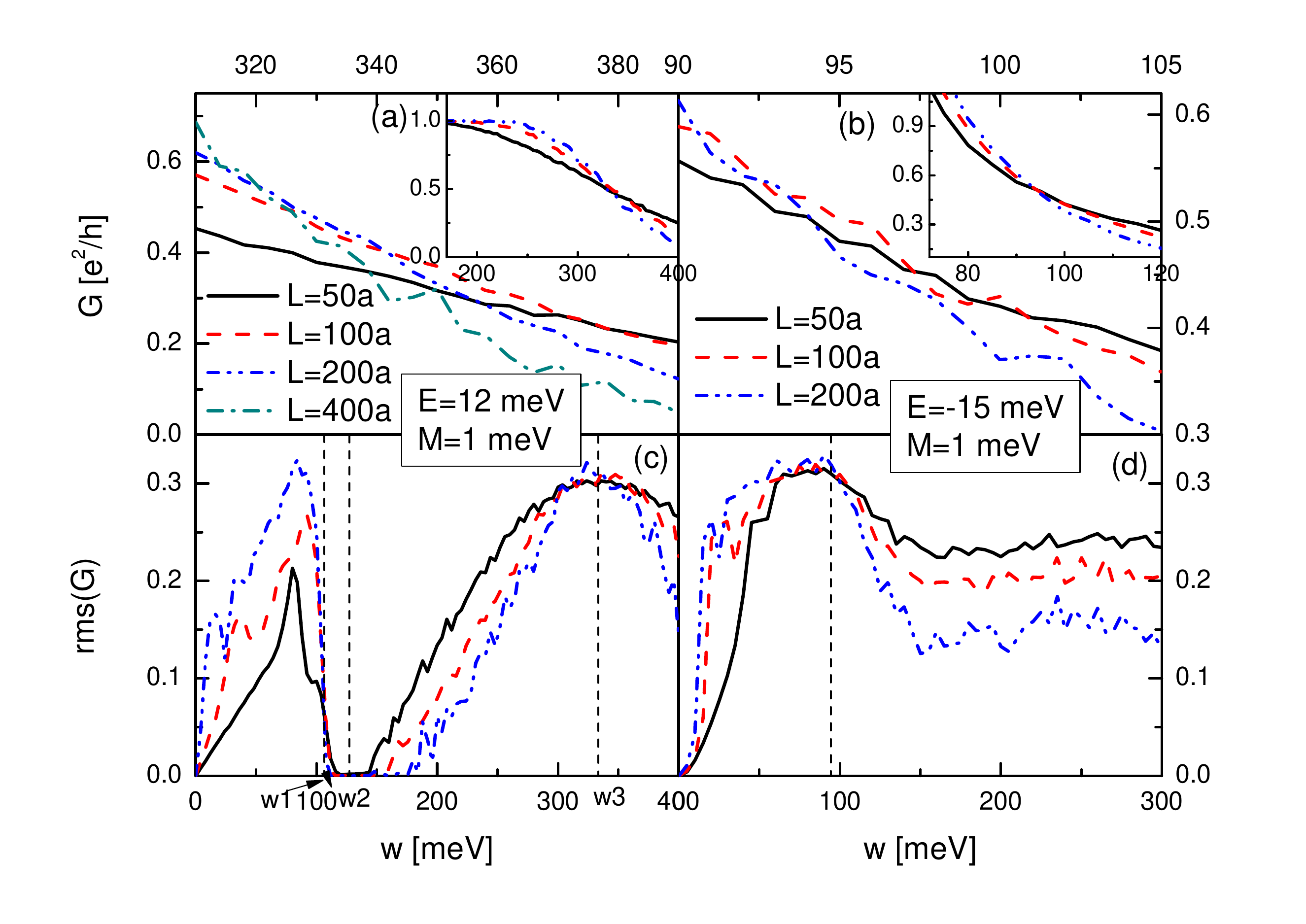}
   \end{center}
   \caption{Anderson transitions to the localized phase. Panes a and c show the transition from the TAI conductance plateau to the insulating phase, while panes b and d show the transition from the bulk conducting phase to the insulating phase.  The insets show the crossing points in a larger perspective. The vertical lines show the phase transition boundaries obtained using sizes $L=50,100,200 a$ (same as Figures ~\ref{fig:E12} and  ~\ref{fig:PhaseDiagram1}).}
   \label{fig:PhaseTransitionCloseups2}
\end{figure}

Figure ~\ref{fig:PhaseTransitionCloseups2} shows the two localization transitions: the quantized-to-localized transition in panes a and c, and the bulk conducting-to-localized transition in panes b and d.  The upper panes show closeups, while the insets show larger scales.  The closeups show that the crossing points are "walking" to weaker disorder because of finite size effects.   At the quantized-to-insulator transition the $L=50,100a$ crossing point is near $W \approx 380$ meV, the $L=100,200a$ crossing poing is around $W \approx 343 $ meV, and  the $L = 200,400a$ crossing point is at $W \approx 320$ meV.   We can expect the true phase transition to be further to the left. In contrast, the bulk-to-insulator transition  at $E=-15$ meV walks only  $\approx 10$ meV between  the $L=50,100a$ and $L=100,200a$ crossing points.  The improved stability at $E = -15 $ meV  may be caused by the larger density of states at that energy.

Figure ~\ref{fig:PhaseTransitionCloseups2}c plots the second moment of the conductance, which forms a broad peak near the quantized-to-localized phase transition.  The position of the peak seems to be less sensitive to finite size effects than the conductance crossing point.  The peak grows more narrow as the system size is increased, suggesting that in the infinite volume limit the second moment is non-zero only at the phase transition. The peak is caused by bulk state delocalization, which allows the bulk states  to connect and destroy the edge states.

Chen et al \cite{Chen11} calculated the logarithmic average of the bulk conductance at system sizes $70a \times 70a, \, 100a \times 100a, \, 130a \times 130a$.  They used periodic boundary conditions and did not include the conductance of edge states.  In the neighborhood of the quantized-insulator transition they found a finite "metallic" region where $\langle \ln G \rangle$ increases, as is typical of a metal.  They concluded that a metallic phase intervenes between the quantized phase and the insulating phase. 
  In our judgement this conclusion is ill-founded.  Our data  near the quantized-insulator transition shows that the average conductance  (including both bulk and edge states) is less than one at all $50 \leq L \leq 400 a$ and converges to a step function.  This  implies that the bulk conductance is also less than one; the bulk does not become metallic.  In fact Chen et al's   $\langle \ln G \rangle$ is always less than zero.  Both their data and our data are consistent with the delocalization of a single bulk channel, which is necessary to destroy the edge states.  It is likely that a careful calculation of the localization lengths associated with several bulk channels would show that only one bulk channel delocalizes at the quantized-insulator phase transition.

 The finite width of the  region where the bulk channel is delocalized may be a finite-size effect, as is suggested by our data on the second moment $rms(G)$ and the fractal dimension $d_2$.   Both of these quantities have broad peaks near the phase transition, and the peaks become progressively more narrow as the system size is increased.  If  one calculated the width of the delocalized region at several system sizes, one should find that this  region shrinks when the system size is increased.  In an infinite system the delocalized region (and the peaks in $rms(G)$ and $ d_2$) should have  zero width and coincide with the quantized-insulator transition.
 
Lastly, Yamakage et al \cite{Yamakage11}  studied systems of size $L = 16, 32, 64a$ and found a region between the quantized and insulating phases where the localization length increases with the system size, but only in a model where the $s_z$ spin component is not conserved.   The model which we are studying does  conserve $s_z$, and Yamakage et al report that it exhibits no intermediate metallic phase.

\subsection{Infinite volume limit of the gapped TAI-I scaling region\label{FateOfTheGap}}

\begin{figure}
\begin{center}
  \includegraphics[width=16cm]{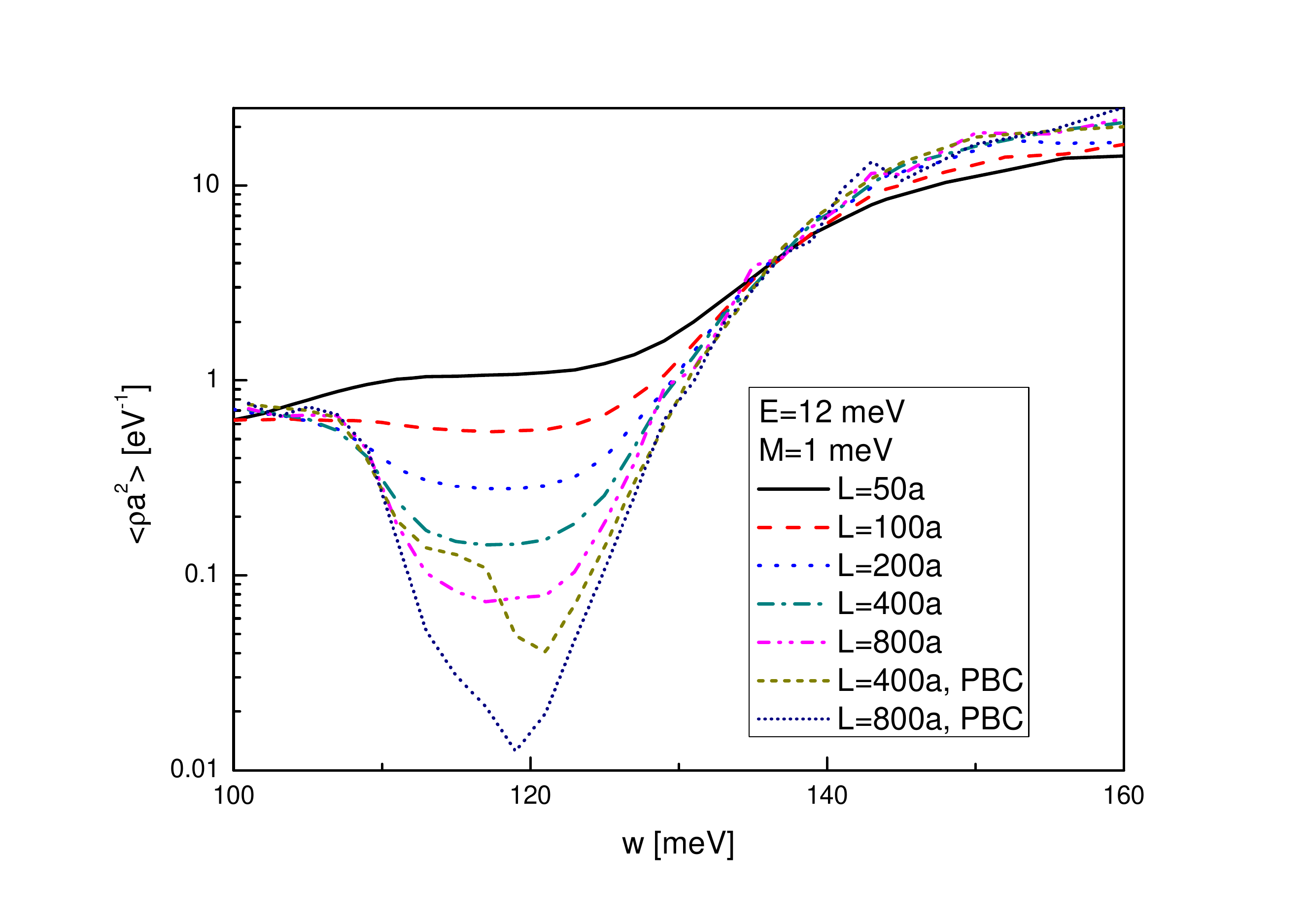}
   \end{center}
   \caption{The TAI-I scaling region at large volumes.  The density of states  converges as the system size increases, starting at the edges of the TAI-I scaling region near $W \approx 110,130$ meV and moving into the center of the scaling region.  This convergence indicates that bulk states dominate the DOS in the infinite volume limit.  The converged DOS agrees well with the "PBC"  lines, which  were obtained with periodic boundary conditions and report the  DOS of bulk states only.  The converged value of the bulk DOS is roughly linear on both sides of the TAI-I  scaling region, which  indicates that the  bulk DOS  is exponentially small, as is  typical of Lifshitz tails.   }
   \label{fig:TAI1InfiniteVolume}
\end{figure}

Figure ~\ref{fig:TAI1InfiniteVolume}  shows the local density of states $\rho$ at system sizes up to $L = 800 a$.  In much of the graph $\rho$ is roughly constant as expected of bulk states; this is the ungapped TAI-II scaling region.   In the interval $\approx 110$  meV $ \leq W \leq \approx 130$  meV $\rho$ decreases as the system size increases from  $L = 50a$ to $L = 100a$ to $L = 200a$.    In most of this interval the $L=50,100,200a$ curves are equally spaced because  edge states obey  $\rho \propto 1/L$.    These signals allow us to distinguish the gapped TAI-I scaling region where there are many more edge states than bulk states.  

As the system size increases, the region where $\rho$ is changing   shrinks,  as does the region of $1/L$ edge scaling. We conclude that $\rho$  converges to a small size-independent value indicating bulk states.  This convergence is confirmed by the $L = 800a \,(PBC)$ curve obtained with periodic boundary conditions, which shows only the bulk  DOS.   (The minima of the PBC curves  suffer from larger errors caused by the very small DOS.)  The linear form of the bulk DOS on both sides of the TAI-I region indicates that the  bulk DOS  is exponentially small. In fact band gaps in disordered systems are always populated by localized intruder states from the bulk bands.  This effect is called a Lifshitz tail, and is known to be exponentially small.  In summary, the DOS is the sum of an edge $1/L$ contribution and an exponentially small bulk contribution from localized Lifshitz tail states.  

Does any observable allow us to distinguish between the TAI-I and TAI-II  scaling regions in  infinite systems?  Since the bulk Lifshitz tails overwhelm the edge states at large system sizes, a scaling analysis of $\rho$ and $\langle d / L \rangle$ will show the gapped TAI-I region disappearing as the size is increased.   However in infinite systems a hole in the bulk DOS will continue to mark the TAI-I region.  The density of states in the TAI-I hole drops exponentially as one moves away from the band edges toward the center of the hole.   Moreover we saw in Figure ~\ref{fig:PhaseTransitionCloseups} that  the TAI-I and TAI-II regions can be distinguished by their transition to the bulk conduction region: the TAI-II  transition manifests scale invariance and a crossing point, while the TAI-I transition exhibits a meeting point and an absence of scale invariance.  Lastly, our calculation of the  $\langle PR \rangle$ with periodic boundary conditions (bulk states only) reveals that at the weak-disorder boundary  the bulk states localize, and that inside the TAI-I region the PR  descends to a plateau  with a very small value $\langle PR \rangle \approx 10 a^2$. The $ \langle PR \rangle$ plateau extends over an interval of about $\approx 20$ meV and includes the boundary between the TAI-I and TAI-II regions.  This bulk localization is responsible for the TAI conductance plateau.  Each of these observables -  the hole in the DOS, the meeting point vs. crossing point, and the very small  $\langle PR \rangle \approx 10 a^2$ - distinguishes between the two TAI scaling regions.

\subsection{Conductance distributions at the phase transitions}

\begin{figure}
\begin{center}
  \includegraphics[width=16cm]{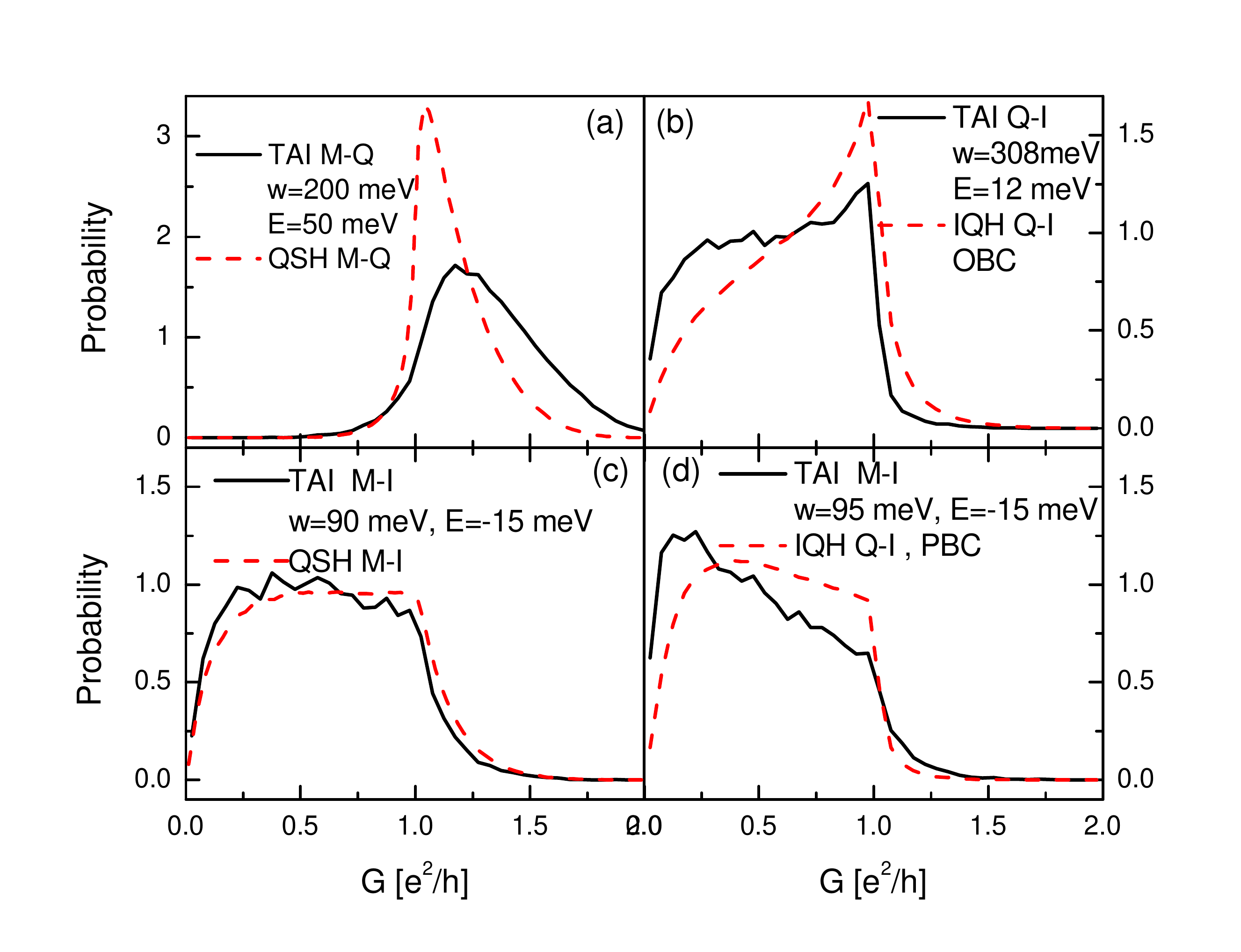}
   \end{center}
   \caption{Conductance distributions at the TAI phase transitions.  Pane a:  the TAI bulk conduction-quantized transition, compared to the Quantum Spin Hall metal-quantized transition.  Pane b: the TAI quantized-insulator transition, compared to the IQH with open boundary conditions.   Pane c: the TAI bulk conduction-insulator transition at $W = 90$ meV, compared to the Quantum Spin Hall metal-insulator transition.  Pane d: the TAI bulk conduction-insulator transition at $W = 95$ meV, compared to the IQH with periodic boundary conditions.  We rescaled Kobayashi et al's QSH data by a factor of $2$ ($P(G) \rightarrow 2 P(G/2)$) because in our own calculations we report only $1/2$ of the total conductance of the $4 \times 4$ Hamiltonian - see the discussion in section \ref{TAIModel}.  The system size is $L = 200 a$ for all of our distributions.} 
   \label{fig:E-15w95condhist}
\end{figure}

Figure ~\ref{fig:E-15w95condhist} shows the conductance probability distribution at four points in the phase diagram.  At disorder-induced phase transitions the conductance distribution is a critical quantity - it converges as the system size is increased, and is independent of microscopic details of  models that belong to the same universality class.  The critical conductance distribution has been determined precisely for both the integer quantum hall system and for the Quantum Spin Hall (QSH) system.    The dashed red lines in Figure ~\ref{fig:E-15w95condhist} were already published in works by Kramer et al \cite{Kramer05} and Kobayashi et al \cite{Kobayashi10}.  The authors   systematically fitted their data to remove finite size effects,  determined precisely the  phase transition positions in the infinite-volume limit, and obtained data sets containing one million conductances.  Here we compare those well known distributions  with the TAI conductance distributions.

 The solid black lines in Figure ~\ref{fig:E-15w95condhist} display our data, which is noisy because our data sets contain only $20,000$ conductances.  This statistical error is dwarfed by our uncertainty about the true infinite-volume limit of the phase transitions.  The conductance distribution can change very quickly as one moves through a phase transition.  Unfortunately a precise determination of the phase transition would have required very thoughtful and systematic fitting of finite size effects, and very large data sets at many system sizes.  Instead we did visual estimates of the crossing points using only a few system sizes.  Therefore our data is valuable mainly for preliminary  suggestions about which universality classes the TAI phase transitions may belong to.
 
 We begin with the quantized-insulator transition, shown in Figure ~\ref{fig:E-15w95condhist}b.  The essential physics here is that the edge states are destroyed while the bulk states are already localized. We compare our data to the integer quantum hall transition because both models are unitary, and because Groth et al \cite{Groth09} suggested that both phase transitions lie in the same universality class.  As we discussed in reference to Figure ~\ref{fig:PhaseTransitionCloseups2}a, this transition has very large finite-size effects and the true transition is likely to lie at  a disorder strength smaller than $\approx 320$ meV, which is the position of the $L=200,400a$ crossing point.  We find that the $W = 308$ meV conductance distribution matches fairly well to the IQH distribution. It is possible that a precise determination of the phase transition would result in a perfect match to the IQH conductance distribution.  As previously mentioned, we measure a fractal exponent $d_2 \approx 1.5$ on this phase boundary, which is consistent with the known IQH  value.
  
 Next we consider the transition from bulk conductance to the ungapped TAI-II scaling region, shown in  Figure ~\ref{fig:E-15w95condhist}a.  The essential physics here is that the bulk states localize while the edge states remain conducting.  Unfortunately there is no metallic phase in the IQH system, so we compare the TAI distribution to the metal-quantized transition in the QSH system.   Despite the fact that the QSH system  belongs to the symplectic not unitary class,  the two distributions have similar shapes, both at $W = 200$ meV (shown) and $W = 205$ meV (not shown).    

Figures ~\ref{fig:E-15w95condhist}c,d show the Anderson transition from bulk conduction to the insulating phase at different values of the disorder $W = 90, 95$ meV.   The essential physics here is that the bulk states localize and that there are no edge states.   Our crossing point analysis (see Figure ~\ref{fig:PhaseTransitionCloseups2}b) along the $E=-15 $ meV line indicates that the infinite volume limit  is at $W < \approx 94$ meV.  Near here the conductance distribution changes very rapidly, perhaps because of the nearby tricritical point where the quantized region begins.  At $W = 90$ meV the  conductance distribution  matches very well to the QSH metal-insulator transition, which is peculiar because the QSH system is symplectic while ours is unitary.    At $W = 95$ meV it is similar to the IQH distribution when edge states have been removed by using periodic boundary conditions.  The similarity is again perplexing, for the IQH transition is a point where the bulk states delocalize briefly, not where they move from extended to localized.   A more precise determination of the Anderson transition is required to determine the true critical conductance distribution.

\section{Conclusion}
In this paper we determined the TAI phase diagram at system sizes $L \leq 280 a$ and studied its evolution in larger systems.  While the TAI conductance plateau is caused by inversion of the bulk bands, it cannot be explained entirely by scattering (CPA) physics and a bulk band gap.  A large portion of the conductance plateau lies in the ungapped TAI-II scaling region.   When $L \leq 280 a$ the gapped TAI-I scaling region is well described by the CPA, but in larger systems its density of states is dominated by  Lifshitz tails of localized bulk states.     We carefully studied the phase transitions in very large systems and excluded the possibility of extra metallic and insulating phases which had been reported in previous studies.   We also reported broad peaks in the eigenstate size and fractal dimension that are centered near the quantized to insulating phase boundary and compared the TAI conductance distributions with known results for IQH and QSH systems.

\textit{Acknowledgements.} We thank Stefan Kettemann, Tomi Ohtsuki, Koji Kobayashi, Carlo Beenakker, Shun-Qing Shen, Rui-Lin Chu, Haiwen Liu, Juntao Song, and Xuele Liu for insightful discussions.

\begin{appendix}

\section{Numerical Errors in Determining the Phase Diagram\label{NumericalErrors}}
Our determination of the crossing points was based on visual comparison of observables at three system sizes, and included neither fitting the data to a finite size scaling function nor a precise mathematical estimate of errors.  There are three error sources:
\begin{enumerate}
\item Finite-size effects.  We used system sizes $L=50,100,200 a$  for determining all fixed-$W$ crossing points (open boxes), and also for  the $ E=-15, \,12, \,35 $ meV data points. At the remaining data points the sizes were $L = 70,\, 140$, and $280 a$. Our phase diagram omits physics at scales greater than $L = 280 a$, and includes finite size effects that are proportional to some  small power of $1/L$.  These effects cause the observed crossing point to be displaced from the position of the true phase transition; the observed crossing point "walks" toward the correct value as the system size is increased.   Figures   ~\ref{fig:PhaseTransitionCloseups}  and    ~\ref{fig:PhaseTransitionCloseups2} and the accompanying discussion report our observations of this effect.  
\item Statistical errors.  Along the upper part of the quantized-insulator transition (filled boxes on line $eg2$) the finite size effects are so large that the statistical errors can be neglected.  Along the lower part of this curve (open boxes on line $eg2$) the finite size effects are much smaller, of the same order as the statistical error.   On the transition from bulk conductance to the ungapped TAI-II scaling region (line $eg1$) the finite size effects are so small that they are overwhelmed by the statistical errors.  In the gapped  TAI-I scaling region the fluctuations in $G$ are essentially nill, so on line $g1$ the statistical error is negligible.  Our estimation of the statistical  errors is based on both calculation of the second moment of the observable and visual observation of  fluctuations in our conductance curves.  Our ensemble size depends on the system size.  For all fixed-$W$ crossing points and also the $ E=-15, \,12, \, 35 $ meV data points, we used $N \geq 3000$ disorder realizations at $L=50a$ and $N \geq 1000, 500$ realizations at $L=100, \,200a$.  At the other fixed-$E$ data points we used $N= 800, 800, 100$ realizations at sizes $L=70, \, 140$, and $280a$.
\item Errors from doping the leads.  As we mentioned in section \ref{TAIModel}, if we leave the leads undoped then they force the conductance to zero at $|E| \leq M = 1 $ meV.  At energies neighboring $|E| \leq M = 1 $ meV there is a finite size effect: at small system sizes the conductance hole is widened.  Therefore we have doped the leads to $E=25 $ meV at all fixed-$W$ data points (the open boxes), and also in Figure ~\ref{fig:E12}.      We examined the doping-induced error by calculating crossing points at $W=200 $ meV and $E=10,\, 12 $ meV both with doped and undoped leads.  Our results indicate that the effect of doping is minimal on line $g1$,  on the fixed-$W$ data points (open boxes), and on Figure ~\ref{fig:E12}.   The doped fixed-$W$ data also satisfies an important consistency check: it reproduces correctly  data points along the lower edge $g2$ of the gapped TAI-I scaling region.   Doping has a much more pronounced effect on line $eg1$, where it shifts the $W=200 $ meV data point from $E=50 $ meV to $E=42 $ meV, but happily all of our data points  on $eg1$ were obtained from undoped data. 
\end{enumerate}

\end{appendix}

\bibliography{Vincent}
\end{document}